\documentclass[12pt,twoside,leqno]{amsart}
\usepackage{mathptmx}
\usepackage{amssymb, amsmath, amsthm, enumerate}
\usepackage{graphicx}    
\usepackage{amsmath,amssymb,latexsym,amsfonts}
\usepackage{cite}
\usepackage{subfig}
\begin{document}
\setcounter{page}{1}
\vspace*{2.0cm}

\title{Reduced order multi switching hybrid synchronization of chaotic systems}
\author[A Khan, D Khattar, N Prajapati] {Ayub Khan$^1$, Dinesh Khattar$^2$, Nitish Prajapati$^{3,*}$}
\date{}
\maketitle

\vspace*{-0.2cm}

\begin{center}
{\footnotesize  $^1$Department of Mathematics, Jamia millia Islamia, Delhi 110025, India\\
$^2$Department of Mathematics, Kirorimal College, University of Delhi, Delhi 110007, India \\
$^3$Department of Mathematics, University of Delhi, Delhi 110007, India \\}
\end{center}
\vskip 2mm

{\footnotesize 

\noindent {\bf Abstract.}  In this article, a new synchronization scheme is presented by combining the concept of reduced-order synchronization with multi-switching synchronization schemes. The presented scheme, reduced-order multi-switching hybrid synchronization, is notable addition to the earlier multi-switching schemes providing enhanced security in applications of secure communication. Based on the Lyapunov stability theory, the active control method is used to design the controllers and derive sufficient condition for achieving reduced-order multi-switching hybrid synchronization between a new hyperchaotic system taken as drive system and Qi chaotic system serving as response system. Numerical simulations are performed in MATLAB using the Runge-Kutta method to verify the effectiveness of the proposed method. The results show the utility and suitability of the active control method for achieving the reduced-order multi-switching hybrid synchronization among dynamical chaotic systems. \vskip 1mm

}

\vskip 6mm

\renewcommand{\thefootnote}{}
\footnotetext{ $^*$Corresponding author
\par
E-mail address: nitishprajapati499@gmail.com
\par
}

\vskip 6mm
\noindent {\bf\large 1. Introduction  }
\vskip 6mm

Synchronization of chaos refers to a process in which two or more chaotic systems (either identical or non-identical) adjust a given property of their motion to a common behaviour. Until 1990, it was considered impractical due to the well-known divergence of trajectories caused by sensitivity of chaotic systems to initial conditions. But the pioneering work of Pecora and Carroll on synchronization of two identical chaotic systems \cite{Ref1} has led the topic to become an interesting area of research. It has been developed and studied comprehensively in the ensuing time. Little over the past two decades various methods for synchronization of chaotic systems have been proposed such as linear and nonlinear feedback synchronization \cite{Ref2,Ref3,Ref4}, adaptive feedback control \cite{Ref5}, active control \cite{Ref6}, optimal control \cite{Ref7}, time delay feedback approach \cite{Ref8}, sliding mode control \cite{Ref9}, backstepping design method \cite{Ref10}, tracking control \cite{Ref11} and so on. Due to fast growing interest in chaos synchronization variety of synchronization types and schemes have also been investigated such as complete synchronization \cite{Ref12}, phase synchronization \cite{Ref13}, anti-phase synchronization \cite{Ref14}, lag synchronization \cite{Ref15}, generalized synchronization \cite{Ref16}, anti-synchronization \cite{Ref17}, projective synchronization \cite{Ref18}, function projective synchronization \cite{Ref19}, hybrid synchronization \cite{Ref20}. 

Recently the problems related to reduced-order synchronization of chaotic systems have attracted the attention of some researchers due to its wide occurrence in biological and social sciences \cite{Ref21,Ref22,Ref23}. The prominent feature of reduced-order synchronization is the synchronization of state variables of the response system with the projections of state variables of drive system where the order of the response system is less than the order of the drive system. Here the number of first-order differential equations is referred to as order. During synchronization all states of the response system will be synchronized. There is increasing interest in the study of chaotic synchronization involving different structures and orders due to its excellent practical applications. For instance, synchronous activity is inevitable in the thalamic and hippocampal neurons network even though the network comprises neurons of different dynamical structures and orders \cite{Ref24}. Another such example is the synchronization that takes place between hearts and lungs where one can observe that circulatory and respiratory systems synchronize with different order \cite{Ref25}. Having potential application in various fields of engineering, reduced-order synchronization have been reported in some recent papers \cite{Ref26,Ref27}. Despite these works, in the overall context of synchronization studies, we can say that the work on reduced-order synchronization forms only a small part of studies on synchronization and much remains unexplored in this area of study. 

To enhance the security of transmission of information via synchronization, multi-switching synchronization was proposed by Ucar et al. \cite{Ref28} based on active control mechanism. The notion behind this method is that in drive-response synchronization, different states of the response system are synchronized with the desired state of the drive system in a multi-switching manner. If at least one state variable of the response system synchronizes with itself in drive system then partially switched synchronization is achieved, and if each state variable of response system synchronizes with a different state variable of drive system then complete switched synchronization occurs. Due to the unpredictability of the switched states this synchronization scheme provides an additional security in secure communication. Inspite of its clear importance to secure communication and chaotic encryption schemes, only a few studies of this kind of synchronization have been reported \cite{Ref29,Ref30,Ref31,Ref32}. To the best of our knowledge there have been no previous studies on reduced-order multi-switching synchronization.

In the phenomenon of hybrid synchronization of chaotic systems, coexistence of complete synchronization and anti-synchronization occurs. Sometimes because of this coexistence this type of synchronization is also referred to as mixed synchronization \cite{Ref33}. Complete synchronization implies that the differences of state variables of synchronized systems with different initial values converge to zero. Anti-synchronization is a phenomenon in which the state variables of synchronized systems with different initial values have the same absolute values but of opposite signs. Consequently, the sum of two signals is expected to converge to zero when anti-synchronization occurs. The coexistence  of complete and anti-synchronization turns out to be very useful in applications of secure communication. Many successful studies have been reported on hybrid synchronization of chaotic systems \cite{Ref34,Ref35,Ref36}.

Motivated by the above studies, in this paper, we present the reduced-order multi-switching hybrid synchronization. We believe that this is the first work which addresses the problem of reduced-order multi-switching hybrid synchronization of chaotic systems. The signal transmitted via such synchronization method will provide strong\-er resistance than most of the usual secure communication schemes based on synchronization when under attack from intruders, thereby ensuring better security when applied in communication applications. Using the active control method we design suitable controllers to achieve the reduced-order multi-switching hybrid synchronization between Qi chaotic system \cite{Ref37} and a new hyperchaotic system designed recently \cite{Ref38} from Liu chaotic system. Synchronization of this new hyper chaotic system has not been investigated earlier. This paper will have a significant contribution to the scientific literature on methods of synchronization for non-linear dynamical systems. The rest of the paper is organised as follows. In Section 2 the description of the chaotic systems is given. In Section 3 reduced-order multi-switching hybrid synchronization between the new hyperchaotic system and Qi chaotic system is presented. The corresponding numerical simulations and results are presented in Section 4. We conclude the paper in Section 5. 

\vskip 6mm
\noindent {\bf\large 2. Problem Formulation and System Description}
\vskip 6mm

\noindent {\bf\large 2.1. Problem Formulation}

Let us consider the chaotic drive system described by
\begin{equation} \label{en1}
\dot{x} = f(x),
\end{equation}
where $x = (x_1, x_2, ..., x_n) \in R^n$ is the state vector of the system (\ref{en1}) and $f: R^n \rightarrow R^n$ is a non-linear continuous vector function. As a response system we consider the chaotic system given by
\begin{equation} \label{en2}
\dot{y} = g(y) + U,
\end{equation}
where $y = (y_1, y_2, ..., y_m) \in R^m$ is the state vector of the system (\ref{en2}), $g: R^m \rightarrow R^m$ is a continuous vector function including non-linear terms and $U = (u_1, u_2, ..., u_m) \in R^m$ is the controller to be designed which synchronizes the state of drive and response system.

When $m < n$ (of course $f \neq g$), that is, the order of response system is less than that of the drive system, then the synchronization is only attained in the reduced order. In particular, reduced order synchronization is the problem of synchronizing a response system with the projection of drive system. Consequently, we can divide the drive system into two parts. The projection part is 
\begin{equation} \label{en3}
\dot{x}_{p} = f_p(x),
\end{equation}
where $x_p = (x_{p_1}, x_{p_2}, ..., x_{p_m}) \in R^m$ and $f_p: R^n \rightarrow R^m$. Remaining part of the system is 
\begin{equation} \label{en4}
\dot{x}_{r} = f_r(x),
\end{equation}
where $x_r \in R^l$, $f_r: R^n \rightarrow R^l$ and order $m, l$ satisfy $m+l=n$. Define the error states between projection (\ref{en3}) of drive system and response system (\ref{en2}) as $e_{ij} = y_j \pm x_{p_i} $ where $i, j$ are indices of the error and $i, j = 1, 2, ..., m$. 
\vskip 2mm
\noindent {\bf Definition 2.1.1. } The drive system (\ref{en1}) and the response system (\ref{en2}) are said to be in reduced order multi-switching synchronization, if there exists appropriate controller $U = (u_1, u_2, ..., u_m)$, such that, 
\begin{equation}
\lim_{t \to \infty} e_{ij} = \lim_{t \to \infty} [ y_j \pm x_{p_i}] = 0,
\end{equation}
 ($i, j = 1, 2, ..., m$ and $i \neq j$ for at least one state variable) which implies that the error dynamic system between the state variables of the projection (\ref{en3}) of drive system and state variables of response system (\ref{en2}) is asymptotically stable.
\vskip 6mm 
\noindent {\bf\large 2.2. System Description}

In \cite{Ref38} a new hyperchaotic system was presented. The 3D Liu chaotic system \cite{Ref39} was altered by joining an additional state $w$ and system structure and parameters were modified to obtain the following hyperchaotic system :
\begin{equation} \label{eq1}
\left\{
\begin{aligned}
\dot{x} & =-a_1x+a_2y,\\
\dot{y} & = a_3x-xz-y+w,\\
\dot{z} & = x^2-a_4(x+z),\\
\dot{w} & = -a_5x,
\end{aligned}
\right.
\end{equation}
where $(x, y, z, w)^T$ is the state vector, and $a_i(i = 1, 2,. . ., 5)$ are positive constant parameters of the system. The introduced state $w$ can be considered as a simple external feedback controller which is connected with the system state $x$. Therefore, from the view of anticontrol of chaos \cite{Ref40}, $w$ can enhance the chaotic dynamical behaviour of the original 3D system. For the parameters $a_1=25$, $a_2=60$, $a_3=15$, $a_4=4$, and $a_5=5$ the system exhibits chaotic behaviour. The 3D phase portraits of the system in different phase spaces with initial values of the hyperchaotic system as ($x(0)=7$, $y(0)=-4$, $z(0)=3$, $w(0)=-1$) are shown in Figure 1.
\begin{figure}[htbp] 
   \centering
    \includegraphics[width=\linewidth,height=3in]{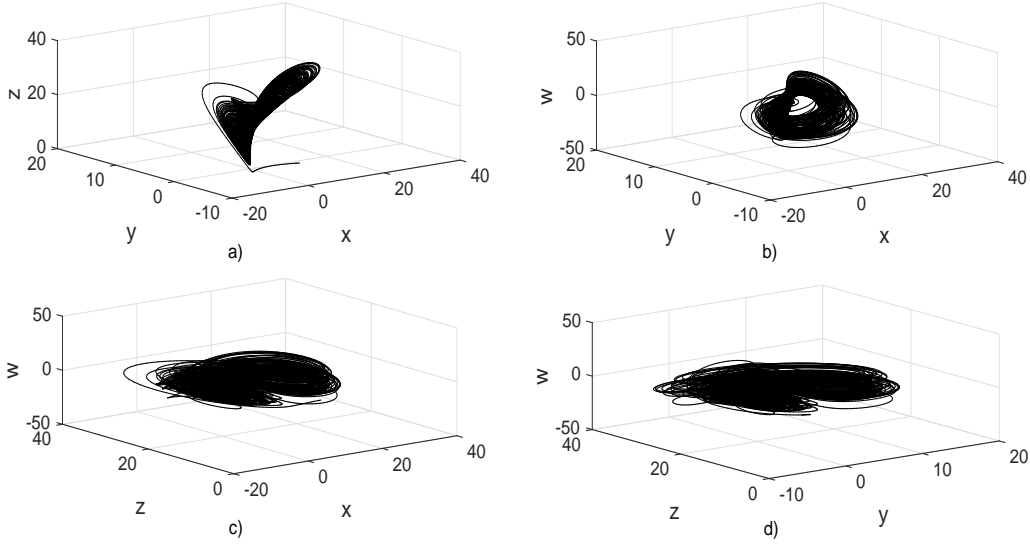}     
    \caption{Phase portraits of new hyperchaotic system in (a) $x-y-z$, (b) $x-y-w$, (c) $x-z-w$, and (d) $y-z-w$ phase space respectively.}
  \end{figure}
  
Qi et al. \cite{Ref37} reported and analysed a new 3D chaotic system in 2005. The system was obtained by adding a cross-product nonlinear term to the first equation of the Lorenz system, forming a new system as follows:
\begin{equation} \label{eq2}
\left\{
\begin{aligned}
\dot{x} &= a(y-x)+yz,\\
\dot{y} &= bx-y-xz,\\
\dot{z} &= -cz+xy,\\
\end{aligned}
\right.
\end{equation}
where notably each equation has one single cross-product term, different from the Lorenz system, Rossler system, Chen system and Lu system, and even Lorenz system family. Here $(x, y, z)^T$ is the state vector, and $a$, $b$, and $c$ are system parameters. For system parameters $a=35$, $b=80$, and $c=8/3$ the system displays chaotic behaviour. With the initial conditions ($x(0)=1$, $y(0)=-6$, $z(0)=5$), the 3D phase portrait of the system with it's 2D phase plane projections is shown in Figure 2.
\begin{figure}[htbp] 
   \centering
    \includegraphics[width=\linewidth,height=3in]{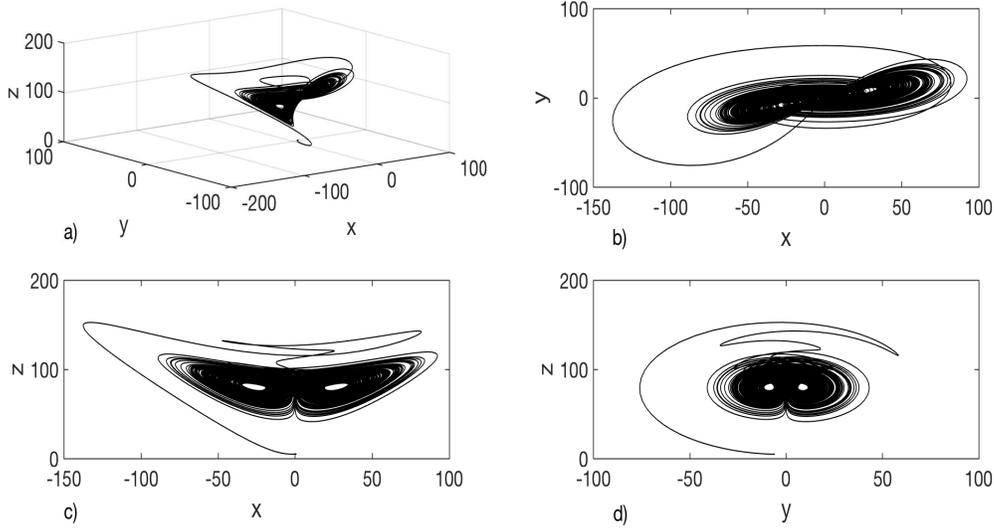}
    \caption{Phase portraits of Qi chaotic system in (a) $x-y-z$ space (b) $x-y$ plane, (c) $x-z$ plane, and (d) $y-z$ plane.}
\end{figure}

\vskip 6mm
\noindent {\bf\large 3. Reduced-order multi-switching hybrid synchronization between the projection of new Hyperchaotic system and Qi chaotic system}
\vskip 6mm

In this section we analyse the reduced-order multi-switching hybrid synchronization between the new hyperchaotic system treated as drive system and Qi chaotic system as response system. To achieve the desired reduced-order synchronization we consider the 3D projection of the drive system (\ref{eq1}) and synchronize it with the response system. We present all the possible switching combinations for the reduced-order 3D projection of hyperchaotic system in $x-y-z$ space and give numerical results for one randomly selected switching hybrid combination from various possibilities. The results for other switching hybrid possibilities are self-explanatory.
\vskip 6mm
\noindent {\bf\large 3.1. Case I} Let us consider the projection of the hyperchaotic system (\ref{eq1}) in $x-y-z$ phase space. We want to synchronize the response system to this projection of hyperchaotic system. The projection in $x-y-z$ phase space of the hyperchaotic system is described by
\begin{equation} \label{eq3}
\left\{
\begin{aligned}
\dot{x_1} & =-a_1x_1+a_2y_1,\\
\dot{y_1} & = a_3x_1-x_1z_1-y_1+w_1,\\
\dot{z_1} & = {x_1}^2-a_4(x_1+z_1),
\end{aligned}
\right.
\end{equation}
and the response system is given by
\begin{equation} \label{eq4}
\left\{
\begin{aligned} 
\dot{x_2} &= a(y_2-x_2)+y_2z_2+u_1,\\ 
\dot{y_2} &= bx_2-y_2-x_2z_2+u_2,\\
\dot{z_2} &= -c_2z_2+x_2y_2+u_3,
\end{aligned}
\right.
\end{equation}
where $u_1$, $u_2$, and $u_3$ are the controllers to be designed in such a manner that the projection (\ref{eq3}) of drive system (\ref{eq1}) and response system (\ref{eq4}) are synchronized. For achieving multi-switching synchronization various switching combinations are possible. Furthermore, since our aim is to achieve multi-switching hybrid synchronization, for any non-specific switching combination, we may choose the states that are to be completely synchronized or anti-synchronized. The following switched hybrid combinations are possible for drive-response systems (\ref{eq3}) and (\ref{eq4}):
\begin{equation*} 
\left\{
\begin{aligned} 
e_{111} &= x_2 \pm x_1,\\ 
e_{112} &= y_2 \pm y_1, \quad \text{Switch 1}\\
e_{113} &= z_2 \pm z_1.
\end{aligned}
\right.
\end{equation*}
\begin{equation*} 
\left\{
\begin{aligned} 
e_{121} &= x_2 \pm x_1,\\ 
e_{122} &= z_2 \pm y_1, \quad \text{Switch 2}\\
e_{123} &= y_2 \pm z_1.
\end{aligned}
\right.
\end{equation*}
\begin{equation*} 
\left\{
\begin{aligned} 
e_{131} &= z_2 \pm x_1,\\ 
e_{132} &= y_2 \pm y_1, \quad \text{Switch 3}\\
e_{133} &= x_2 \pm z_1.
\end{aligned}
\right.
\end{equation*}
\begin{equation*} 
\left\{
\begin{aligned} 
e_{141} &= y_2 \pm x_1,\\ 
e_{142} &= x_2 \pm y_1, \quad \text{Switch 4}\\
e_{143} &= z_2 \pm z_1.
\end{aligned}
\right.
\end{equation*}
\begin{equation*}
\left\{
\begin{aligned} 
e_{151} &= y_2 \pm x_1,\\ 
e_{152} &= z_2 \pm y_1, \quad \text{Switch 5}\\
e_{153} &= x_2 \pm z_1.
\end{aligned}
\right.
\end{equation*}
\begin{equation*} 
\left\{
\begin{aligned} 
e_{161} &= z_2 \pm x_1,\\ 
e_{162} &= x_2 \pm y_1, \quad \text{Switch 6}\\
e_{163} &= y_2 \pm z_1.
\end{aligned}
\right.
\end{equation*}

Switch $i$, ($i=2, 3, 4$) is an example of partial switching where at least one state variable synchronizes to itself and rest are synchronized in some cross combination manner. Switch $j$, ($j=5, 6$) characterises complete switching in which each state variable is synchronized with some different state variable. To investigate the multi-switching hybrid synchronization between these systems let us define the error states as:
\begin{equation} \label{eq5}
\left \{ 
\begin{aligned}
e_{151} &= y_2-x_1,\\ 
e_{152} &= z_2-y_1, \\
e_{153} &=x_2+z_1. 
\end{aligned}
\right.
\end{equation}
It is evident from the error states that different states of slave system are synchronized with desired state of master system in complete switched manner. Moreover it is also clear that there is co-existence of complete synchronization and anti-synchronization among the error states so that hybrid synchronization can be achieved.

The corresponding error dynamics is obtained as
\begin{equation} \label{eq6}
\left\{
\begin{aligned} 
\dot{e}_{151} =& -(1+a_1)e_{151}+bx_2-x_2z_2-a_2y_1-x_1+a_1y_2+u_2,\\ 
\dot{e}_{152} =& -(c+1)e_{152}+x_2y_2-a_3x_1+x_1z_1-w_1\\
                        &-cy_1+z_2+u_3,\\
\dot{e}_{153} =& -(a+a_4)e_{153}+ay_2+y_2z_2+(x_1-a_4)x_1+az_1\\
                       &+a_4x_2+u_1.
\end{aligned}
\right.
\end{equation}
\vskip 2mm
\noindent {\bf Theorem 3.1.1.} If the control functions $u_1$, $u_2$, and $u_3$ are designed as
\begin{equation} \label{eq7}
\left\{
\begin{aligned} 
{u_1} &= -(a+z_2)y_2+(a_4-x_1)x_1-az_1-a_4x_2+V_1,\\ 
{u_2} &= (z_2-b)x_2+a_2y_1+x_1-a_1y_2+V_2,\\
{u_3} &= (a_3-z_1)x_1-x_2y_2+w_1+cy_1-z_2+V_3,
\end{aligned}
\right.
\end{equation}
where $V_1$, $V_2$, and $V_3$ are the functions of the error states $E_1$, $E_2$, and $E_3$, where $E_1=e_{151}$, $E_2=e_{152}$, and $E_3=e_{153}$, chosen such that the error dynamical system (\ref{eq6}) becomes stable, then the drive system (\ref{eq3}) and response system (\ref{eq4}) will achieve the reduced-order complete-switched hybrid synchronization.

\vskip 2mm
\noindent{\bf Proof. }The error dynamical system is given by (\ref{eq6}). Now using the controllers defined by (\ref{eq7}) in (\ref{eq6}) we get
\begin{equation} \label{eq8}
\left\{
\begin{aligned} 
\dot{E_1} &= -(1+a_1)E_1+V_2,\\ 
\dot{E_2} &= -(c+1)E_2+V_3,\\
\dot{E_3} &= -(a+a_4)E_3+V_1.
\end{aligned}
\right.
\end{equation}
We choose $V_1$, $V_2$, and $V_3$ in such a manner that the closed loop system (\ref{eq8}) becomes stable. As long as these feedback stabilize the system, the errors $E_1$, $E_2$, and $E_3$ will asymptotically converge to zero i.e. zero solution of system (\ref{eq8}) will be asymptotically stable. Let us choose
\begin{equation*} 
\begin{bmatrix}
V_1\\
V_2\\
V_3
\end{bmatrix} =A
\begin{bmatrix}
E_1\\
E_2\\
E_3
\end{bmatrix},
\end{equation*}
where $A$ is $3 \times 3$ constant matrix whose elements are chosen such that $V_1$, $V_2$, and $V_3$ make (\ref{eq8}) stable. Various choices of $A$ are possible. We choose a particular form of matrix $A$ given by
\begin{equation*}
A=
\begin{bmatrix}
0 & 0 & a\\
a_1 & 0 & 0\\
0 & c & 0
\end{bmatrix}.
\end{equation*}
Thus \begin{equation*} 
\begin{bmatrix}
V_1\\
V_2\\
V_3
\end{bmatrix} =
\begin{bmatrix}
aE_3\\
a_1E_1\\
cE_2
\end{bmatrix}.
\end{equation*}
The error system (\ref{eq8}) with these values of $V_1$, $V_2$, and $V_3$ becomes
\begin{equation} \label{eq9}
\left\{
\begin{aligned} 
\dot{E_1} &= -E_1,\\ 
\dot{E_2} &= -E_2,\\
\dot{E_3} &= -a_4E_3.
\end{aligned}
\right.
\end{equation}
To show that the closed loop system (\ref{eq8}) and equivalently system (\ref{eq9}) is stable under this choice of $V_1$, $V_2$, and $V_3$, let us construct the Lyapunov function $V$ as follows:
\begin{equation} \label{eq10}
V(t)=\frac{1}{2}(k_1E_1^2+k_2E_2^2+k_3E_3^2),
\end{equation}
where $k_1$, $k_2$, and $k_3$ are positive numbers. This function is positive definite in $\mathbb{R}^3$. Taking the time derivative of $V$ along the trajectory of error dynamical system we obtain
\begin{align*}
\frac{dV}{dt} &= k_1E_1\dot{E_1}+k_2E_2\dot{E_2}+k_3E_3\dot{E_3},\\ 
                    &= k_1E_1(-E_1)+k_2E_2(-E_2)+k_3E_3(-a_4E_3),\\ 
                    &= -k_1E_1^2-k_2E_2^2-k_3a_4E_3^2. 
\end{align*}
As $V$ is positive definite function and $\dot{V}$ is negative definite function, consequently, according to Lyapunov stability theorem (\ref{eq8}) becomes stable. Equivalently we can say that the zero solution of the system (\ref{eq9}) is asymptotically stable. As a result, the drive system (\ref{eq3}) and response system (\ref{eq4}) achieve the reduced-order multi-switching hybrid synchronization. This completes the proof. 

\noindent {\bf\large 3.2. Case-II, Case III, and Case IV} The results for the projection of hyperchaotic system in $x-y-w$, $x-z-w$, and $y-z-w$ phase planes are obtained similarly as above.

\vskip 6mm
\noindent {\bf\large 4. Numerical simulations and results}
\vskip 6mm
\begin{figure}[tbp] 
   \centering
   \includegraphics[width=\linewidth,height=3in]{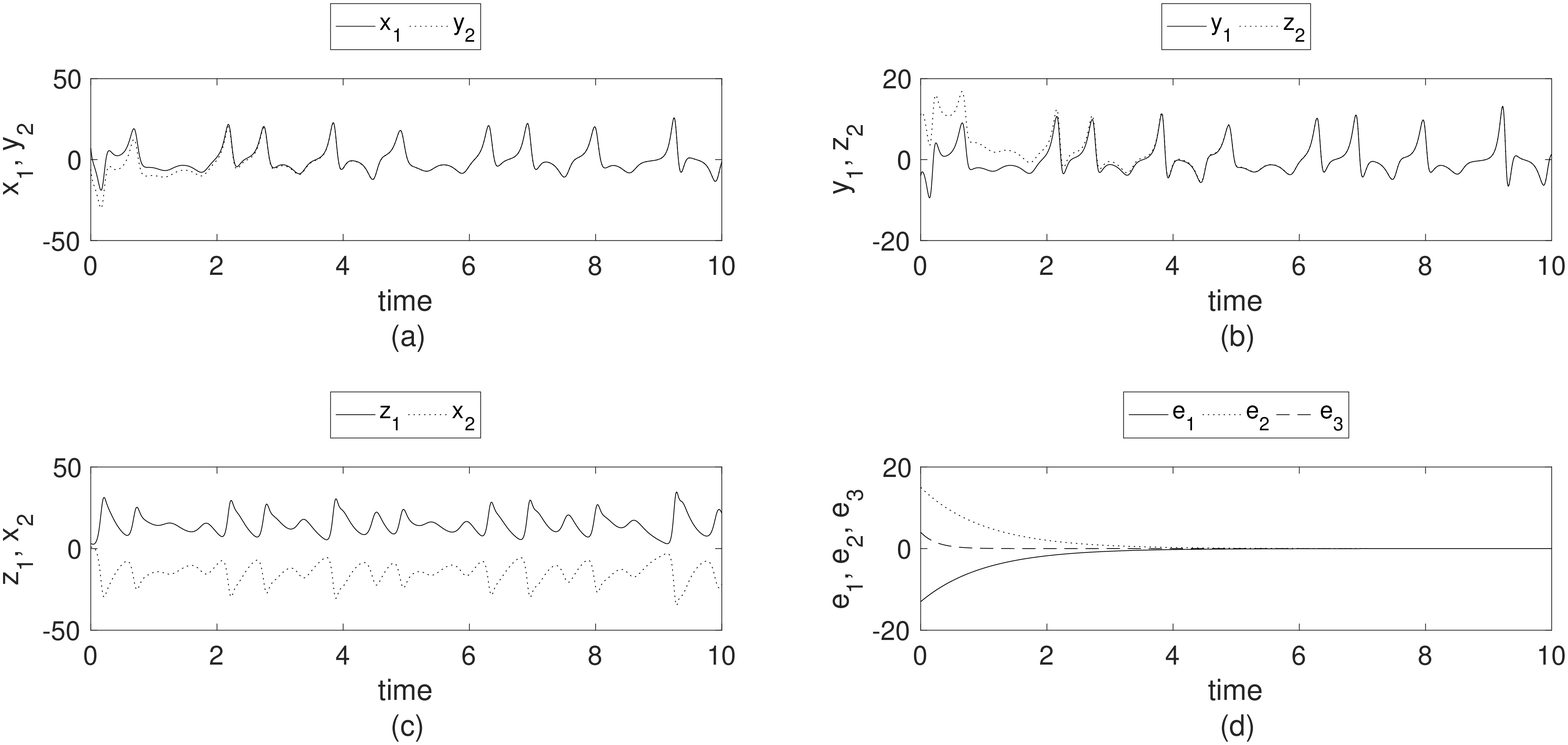} 
   \caption{Multi-switching hybrid synchronization between: (a) $x_1-y_2$, (b) $y_1-z_2$, (c) $z_1-x_2$ and (d) the evolution of the error functions $e_1, e_2$ and $e_3$}
\end{figure}

All simulations are carried out using MATLAB. The fourth-order Runge-Kutta method is used to solve the systems of differential equations. The system parameters for hyper chaotic system and Qi chaotic system are taken as ($a_1=25$, $a_2=60$, $a_3=15$, $a_4=4$, $a_5=5$) and ($a=35$, $b=80$, $c=8/3$) respectively. In Case-I the initial values for drive and response systems (\ref{eq3}) and (\ref{eq4}) are taken as ($x_1(0)=7$, $y_1(0)=-4$, $z_1(0)=3$) and ($x_2(0)=1$, $y_2(0)=-6$, $z_2(0)=11$). Our aim is to achieve multi-switched hybrid synchronization among the state variables. To achieve the desired result we define errors in this case as given by (\ref{eq5}). Observe that the errors are defined such that the state variables $y_2$, and $z_2$ of the response system are completely synchronized with the state variables $x_1$, and $y_1$ of the drive system respectively and variable $x_2$ becomes anti-synchronized with variable $z_1$. This error state corresponds to Switch $5$ of Case-I, which gives us complete-switched hybrid synchronization among the state variables. The initial error values that we obtain from the error states (\ref{eq5}) are ($E_1(0)=-13$, $E_2(0)=15$, $E_3(0)=4$). From Figure 3 we can see that after using the controllers defined in (\ref{eq7}) the error dynamics converges to zero and the drive system (\ref{eq3}) and response system (\ref{eq4}) achieve the desired reduced-order multi-switching hybrid synchronization amongst their state variables.

\vskip 6mm
\noindent {\bf\large 5. Conclusion}
\vskip 6mm

We have introduced and analysed a new form of synchronization called reduced-order multi-switching hybrid synchronization. In this scheme, the different state variables of the response system are synchronized with the desired state variables of projection of the drive system. The presented synchronization scheme can be used to improve the security of information transmission. As there are several possible synchronization combinations, it would be very difficult for the intruder to determine the combination in which synchronization would occur. Using active control we have been successful in designing the controllers to achieve reduced-order multi-switching hybrid synchronization between a new hyperchaotic system treated as drive system and Qi chaotic system as response system. The synchronization of this new hyperchaotic system has not been reported earlier. This paper leads the way in synchronization studies of this new system. Numerical simulations using Runge Kutta method have been performed to verify the analytical results. The reduced-order multi-switching synchronization scheme presented in this paper is attractive method of synchronization as it gives more switching options for constructing the error states, largely possible because of the hyperchaotic nature of drive system. Finally, this synchronization scheme may be applied to other chaotic systems as well and the results obtained here present new directions in the study of various kinds of chaotic synchronizations.

\vskip 6mm
\section*{\bf Acknowledgements}
The work of the third author is supported by the Junior Research Fellowship of Council of Scientific and Industrial Research, India(Grant no. 09/045(1319)/2014-EMR-I).

\end{document}